\documentclass[aps,preprint,amssymb]{revtex4}

\usepackage{epsfig}

\begin{document}
\renewcommand{\thefootnote}{\fnsymbol{footnote}} 
\renewcommand{\theequation}{\arabic{section}.\arabic{equation}}

\title{Inner Shell Definition and Absolute Hydration Free Energy of K$^+$(aq) 
on the Basis of Quasi-chemical Theory and {\it Ab Initio} Molecular Dynamics} 

\author{Susan B. Rempe,$^a$ D. Asthagiri$^b$ and Lawrence R. Pratt$^b$}
\email[]{E-mail: lrp@lanl.gov}
%\thanks{}

\affiliation{$^a$ Sandia National Laboratories, Albuquerque, NM 87185, USA\\
$^b$ Theoretical Division, Los Alamos National Laboratory, Los 
Alamos NM 87545, USA}

\date{\today}

\begin{abstract}
\noindent 
K$^+$(aq) ion is an integral component of many cellular processes,
amongst which the most important, perhaps, is its role in transmitting
electrical impulses along the nerve. Understanding its hydration
structure and thermodynamics is crucial in dissecting its role in such
processes. Here we address these questions using both the statistical
mechanical quasi-chemical theory of solutions and {\em ab initio\/}
molecular dynamics simulations. Simulations predict an interesting
hydration structure for K$^+$(aq): the population of about six (6) water
molecules  within the initial minimum of the observed $g_{KO}(r)$ at
infinite dilution involves four (4) inner-most molecules that the
quasi-chemical theory suggests should be taken as the theoretical inner
shell. The contribution of the 5$^{\underline{th}}$ and
6$^{\underline{th}}$ closest water molecules is observable as a distinct
shoulder on the principal maximum of the $g_{KO}(r)$. The quasi-chemical
estimate of solvation free energy for the neutral pair KOH is also in
good agreement with experiments.
\end{abstract}

\maketitle

\section{Introduction}
Many of the most characteristic molecular processes in aqueous
solutions, including biophysical processes, depend upon electrolytic
properties of these phases. Molecular scale understanding of the
statistical thermodynamics of ionic species in water is important for
understanding those processes. Both theory and simulation of these
systems has become much more revealing in recent years. As successive
cases of common ions are addressed with more %revealing 
informative tools \cite{DA:03},
idiosyncratic differences between chemically similar ions, for example,
the alkali metals, come to the forefront. This paper considers
K$^+$(aq) in detail, and studies an intrinsic difference from the cases
of Li$^+$(aq) and Na$^+$(aq); specifically, the definition on the basis
of theory and simulation observation of an inner hydration shell.

A satisfactory identification of an inner shell makes molecular theory
and other subsequent considerations much simpler. For Li$^+$(aq)
\cite{lrp:jacs00,lrp:ES99,LyubartsevAP:HydLiA,LoefflerHH:Thehsl,%
EgorovAV:Temace,LoefflerHH:LitihQ} and Na$^+$(aq)
\cite{WhiteJA:ThesNw,lrp:fpe01}, such an inner shell is obvious from the
most primitive observation, the radial distribution of water (oxygens)
conditional on the atomic ion. For K$^+$(aq) that identification is less
trivial, but that is the goal of this note. This is likely to be an
issue of broader relevance to other ions in solution, for example,
HO$^-$(aq) \cite{lrp:HOQCA,lrp:HO02}. 

The more revealing tools hinted at above are {\em ab initio\/} molecular
dynamics (AIMD) and the quasi-chemical theory of solutions. K$^+$(aq) at high
dilution has been the subject of a preliminary AIMD calculation
\cite{RamaniahLM:Abims}.  In contrast, a recent simulation study focused
on infinite dilution hydration free energies utilizing classical force field models
\cite{HerceDH:Calicf}. Concentrated KOH aqueous solutions have been
studied by Car-Parrinello {\it ab initio} molecular dynamics 
methods, too \cite{Klein:jacs02,Klein:jpca02}. A more
expansive molecular dynamics study of KF aqueous solution on the basis
of a classical model force field is presented in \cite{LaudernetY:Amdc}.

\section{{\it Ab Initio} Molecular Dynamics}

The K$^+$(aq) model system consisted of 32 water molecules and one K$^+$
ion contained in a cubic box of length 9.865~\AA\ and subjected to
periodic boundary conditions. With this box volume, the water density in
the system matched the experimental density of bulk liquid water at
standard conditions, with no alterations made for the presence of the
ion.  A structure obtained from a well-equilibrated system used in a
previous study of Na$^+$(aq) \cite{lrp:fpe01} provided a template for
generating initial configurations for the current investigation. The
sodium ion and 8 nearest water molecules were removed from the former
system and replaced with a cluster composed of 8 water molecules
surrounding a K$^+$ ion. The geometry of the inserted K$^+$ cluster was
first optimized in a separate {\it ab initio} calculation and, when
placed in the simulation box, the cluster became 10-fold coordinated due
to the proximity of two other water molecules in the box.  

It is the
dynamical evolution of this initial configuration, determined within an
{\it ab initio} framework,  that is presented and analyzed here.
Alternative initial configurations generated in an analogous manner
using 6 and 9 water molecules were used to confirm that the results were
independent of initial conditions, but will not be addressed further.

{\it Ab initio} molecular dynamics simulations based on the PW91
gradient-corrected electron density functional description of the
electronic structure and interatomic forces were performed on the K$^+$(aq)
system using the VASP program \cite{kresse:prb93, kresse:prb96}.
%In previous work\cite{Li, Na, Be...}, this functional was shown to provide a reasonable description
%of the aqueous ion system.  
Valence electrons, which included the semi-core 3{\it p} states of K,
were treated explicitly, while Vanderbilt \cite{vanderbilt,kresse:jpcm94}
ultrasoft pseudopotentials described core-valence interactions for all
atoms.  The valence orbitals were expanded in plane waves with a kinetic
energy cut-off of 29.1 Ry, and the equations of motion were integrated in
time steps of 1.0 fs for a total simulation time of 20 ps.

\begin{figure} 
\begin{center} 
\includegraphics[width=6.5in]{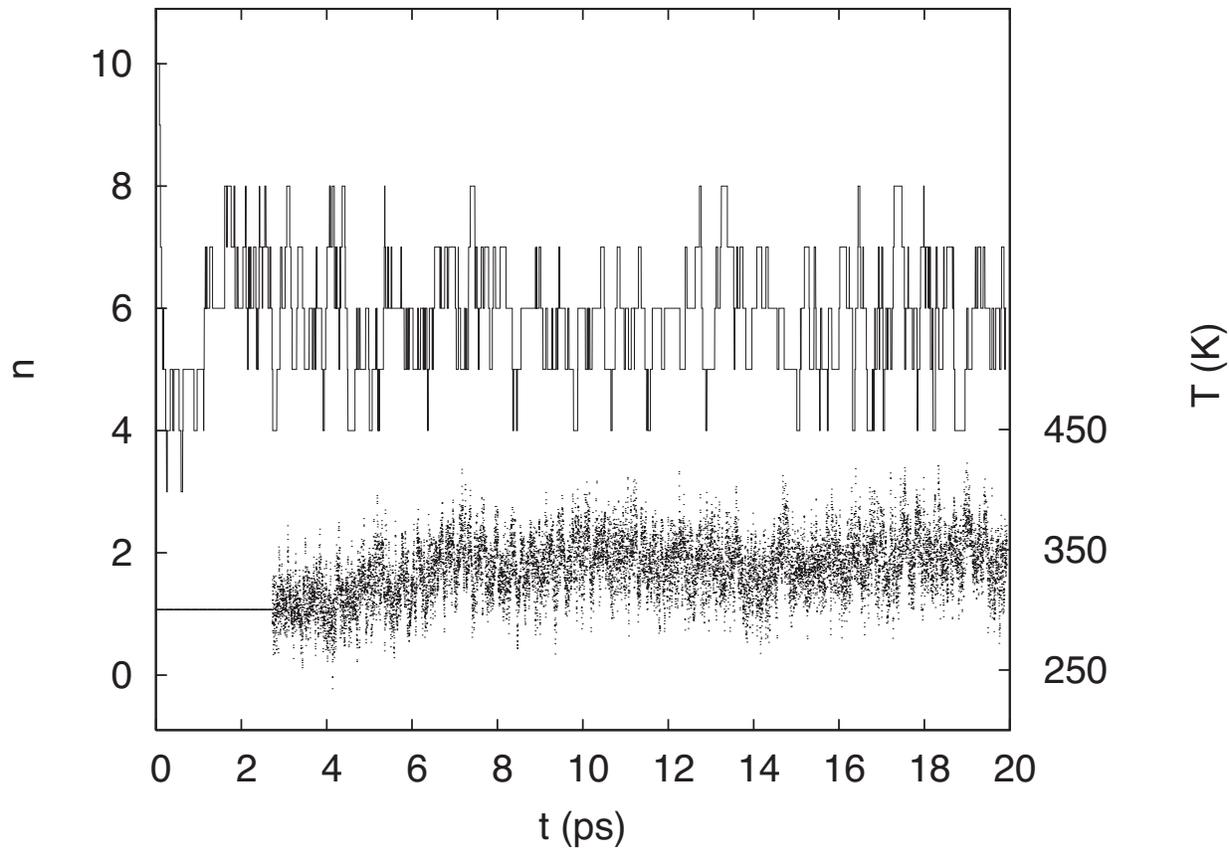} 
\end{center} 
\caption{The upper trace (left axis) follows the number of water
molecules within r=3.50~\AA\ of the K$^+$ ion at each time step in the
dynamical simulation.  The radius defines the inner shell of
coordinating water molecules, as determined by the first minimum in the
radial density distribution analysis (Fig.~\ref{fg:gk}). The lower trace
(right axis) records the instantaneous system temperature.  Only the
last 10~ps of this record was used in the structural analysis. }\label{bonds} 
\end{figure}

For the first 3 ps of the AIMD simulations, a constant temperature of
300~K is maintained by scaling the velocities.  During this time, the
K$^+$ immediately sheds half of its 10 coordinating water molecules by
expanding into a more relaxed geometry with fewer water ligands, see
Fig.~\ref{bonds}. When the temperature constraint is removed, the
system heats up a little and the number of coordinating water molecules
fluctuates between four (4) and eight (8). 
Note that it takes more than 4 ps for the system temperature to stabilize around
a new average value.  The structural analyses presented here utilize the simulation
data collected between 10 and 20 ps, for which the system temperature averages
340$\pm$23~K.

\section{Quasi-chemical Theory} 

The discussion of the quasi-chemical approach \cite{lrp:apc02}, and its
application to ion-solvation problems (see, for example
\cite{lrp:ES99,lrp:ionsjcp03}) has been considered in detail before.
Briefly, the region around the solute of interest is partitioned into
inner and outer shell domains. For K$^+$(aq), the inner shell comprises
the water molecules directly coordinated with the ion. This domain is
treated quantum mechanically. The outer-shell contribution has been
assessed using both a dielectric continuum model and classical force
field simulations (see \cite{lrp:ionsjcp03}).
% How about noting scaled particle theory and perturbation estimations as in Krypton paper
The theory permits a variational check of the inner-outer partition
\cite{lrp:fpe01,lrp:h2oAIMD03}, but this aspect has not been pursued
here.

The inner shell reactions are: 
\begin{eqnarray*} 
\mathrm{K^+ + n H_2O \rightleftharpoons K[H_2O]_n{}^+} 
\end{eqnarray*} 
The free energy change for these reactions were calculated using the
Gaussian programs \cite{gaussian}. The $\mathrm{K\cdot[H_2O]_n{}^+}$ (n
= 0$\ldots$ 8) clusters were geometry optimized in the gas phase using
the B3LYP hybrid density functional \cite{b3lyp} and the 6-31+G(d,p)
basis set. Frequency calculations confirmed a true minimum, and the zero
point energies were computed at the same level of theory. Single point
energies were calculated with the 6-311+G(2d,p) basis.

For estimating the outer shell contribution, the ChelpG method
\cite{breneman:jcc90} was used to obtain partial atomic charges. Then
the hydration free energies of the clusters were calculated using a
dielectric continuum model \cite{lenhoff:jcc90}, with the radii set
developed by Stefanovich {\it et al.} \cite{Stefanovich:cpl95} and
surface tessera generated according to Ref.~\cite{sanner}. With this
information and the binding free energies for the chemical reactions, a
primitive quasi-chemical approximation to the excess chemical potential
of K$^+$(aq) in water is:
\begin{eqnarray} 
\beta \mu_{\mathrm{K}^+(aq)}^{ex} &\approx& - \ln \left\lbrack 
\sum_{n\ge 0} \tilde{K}_n \rho_{\mathrm{H}_2\mathrm{O} }{}^n \right \rbrack 
\label{eq:regrouped} 
\end{eqnarray} 
where $\tilde{K}_n=K_n^{(0)}\exp\left[{-\beta
\left(\mu_{\mathrm{K}(\mathrm{H}_2\mathrm{O})_n{}^+}^{ex}-n
\mu_{\mathrm{H}_2\mathrm{O} }^{ex}\right)}\right]$. $K_n^{(0)}$ is the
equilibrium constant for the reaction in an ideal gas state, $n$ is the
hydration number of the most stable inner shell cluster, and
$\mathrm{\beta=1/k_\mathrm{B}T}$. The density factor
$\mathrm{\rho_{H_2O}}$ appearing in Eq.~\ref{eq:regrouped} reflects the
actual density of liquid water and its effect is included by the
replacement contribution of $-n k_\mathrm{B}T \ln
(\rho_\mathrm{H_2O}/\rho_0)$ = -$n k_\mathrm{B} T \ln (1354) $, where
$\rho_\mathrm{H_2O} = 1~$gm/cm$^3$, and $\rho_0 = 1\,\mathrm{atm}/RT$.
(Discussions on standard states and this replacement contribution can be
found in Pratt and Rempe \cite{lrp:ES99} and Grabowski {\it et
al.}~\cite{lrp:jpca02}.) Note that it is in forming
Eq.~\ref{eq:regrouped} that approximations enter the theory, but all
these approximations are available for scrutiny and improvement.
%\cite{lrp:jpca02}.

Following this procedure, the most probable inner-shell coordination
number was found to be four (4). Based on this identification, we
pursued more refined molecular dynamics calculations to assess the
outer-shell contribution using the $\mathrm{K[H_2O]_4{}^+}$ cluster. The
oxygen and hydrogen atoms of the cluster were assigned the van~der~Waals
parameters of the classical SPC/E \cite{spce} water model. The ChelpG
charges were retained for the cluster. This cluster was placed in a bath
of 306 SPC/E water molecules and its hydration free energy calculated in
the same manner as described in \cite{lrp:ionsjcp03}. In contrast to
this earlier study, only one box size was used. This was founded on the
earlier observation that using different box dimensions did not change
the hydration free energy much, indicating that ion(aq) finite size
effects are modest.  More expansive discussions of outer shell contributions
can be found in \cite{lrp:ES99,lrp:apc02,lrp:h2oAIMD03,Ashbaugh:krcpl03}

\section{Results} 

\subsection{{\it Ab initio} molecular dynamics} 

Fig.~\ref{fg:gk} shows the radial density distribution of the water
oxygen atoms around the ion, obtained by analyzing the last 10 ps of the
simulation using a bin width of 0.1~\AA.  The density of water peaks at
only a little over 2.5 times the density of bulk water and the first
coordination shell is characterized by a broad first peak spanning over
1~\AA\ in width with a shallow ill-defined minimum, in contrast to the
much narrower and better defined distributions of water around the
smaller alkali metal cations, Li$^+$ and Na$^+$ \cite{DA:03}.  The area
under the first peak indicates that the mean number of 5.9$\pm$0.9 water
molecules coordinate the ion within a distance of r=3.50~\AA. Of more
interest is the secondary structure appearing in the first peak,
indicating a composite nature of the water distributed around potassium
ion.

The analysis of Fig.~\ref{fg:gk} suggests that the first coordination
shell is actually composed of two sub populations of water molecules,
with one subset occupying the region closest to the ion and the other
subset situated at a slightly larger distance. In Fig.\ref{fg:gk},
contributions to the radial density distribution of the four (4) water
molecules nearest the K$^+$ ion and the next two (2) water molecules
have been separated out from the full distribution. Two overlapping, but
distinct populations of water molecules are apparent. Note that for $r$
= 3~\AA\, where a shoulder appears on the far slope of the principal
maximum in g(r), the mean occupancy is about four (4). This contrasts to
the mean occupancy of six (6) in the inner hydration shell given by the
conventional definition above, where the first minimum in g(r)
establishes identification of the inner population.

In the analysis presented next, quasi-chemical theory establishes four
(4) as the most probable inner occupancy, in agreement with the first
subpopulation identified in the preceding analysis of the dynamical
data, and on that basis a quasi-chemical analysis provides a good
estimate of the absolute hydration free energy.

%total simulation time was about 20~ps, of which the last 10~ps have been
%used in computing the distribution functions. A bin-width of 0.1~{\AA}
%was used in constructing $\mathrm{g_{KO}}$.

Previous simulation results agree qualitatively with Fig.~\ref{fg:gk},
but differences in resolution \cite{RamaniahLM:Abims}, or concentration
\cite{Klein:jacs02,Klein:jpca02,LaudernetY:Amdc} preclude a more
detailed comparison.

\subsection{Quasi-chemical calculations} 

The hydration free energy of K$^+$, using the classical molecular
dynamics calculation for the outer-shell contribution, is
$-70.5\pm2.1$~kcal/mole for transfer of the solute from 1~mol (ideal
gas) to 1~mol (ideally diluted solute) solution. The experimental values
are suitably adjusted to account for this choice of standard state. As
found before \cite{lrp:ionsjcp03}, the agreement between our absolute
hydration free energies and the values of Coe and coworkers
\cite{coe:jpca98} ($-86$~kcal/mole) is poor. The sign and magnitude of
this discrepancy is in line with discrepancies already identified for
H$^+$, Li$^+$, and Na$^+$ %\cite{lrp:ionsjcp03}. As discussed earlier
%\cite{lrp:ionsjcp03}, this discrepancy suggests that 
and further validates our suggestion that the absolute
hydration free energies \cite{lrp:ionsjcp03} estimated by Coe and coworkers %\cite{coe:jpca98}
contain a negative contribution from the  potential of the phase.

Nevertheless, solvation free energies of neutral combinations, such as
KOH, are thermodynamically unambiguous and can be accessed
experimentally. With the hydration free energy of HO$^-$,
$-123.8$~kcal/mole, computed earlier \cite{lrp:ionsjcp03} using the
SPC/E model for the outer-shell contribution, the hydration free energy
of the neutral combination KOH is $-194.3\pm2.1$. This is in good
agreement with the experimental value  of
$-191$~kcal/mole by Coe {\it et al.}, adjusted for our choice of
standard states.

We note that for the hydration of HO$^-$(aq), the quasi-component was
the tri-hydrated state $\mathrm{HO[H_2O]_3{}^-}$. This choice was
confirmed by both {\it ab initio} molecular dynamics \cite{lrp:HO02} and
quasi-chemical calculations \cite{lrp:HOQCA}.

\section{Conclusions} 

Primitive quasi-chemical theory identifies four (4) as the most probable
water/oxygen  occupancy of a chemically defined inner shell
for the K$^+$(aq) at infinite dilution. On
this basis, that quasi-chemical theory gives a good estimate of the
absolute hydration free energy of K$^+$(aq). These results are
consistent with AIMD observations, but the inner shell is less clearly
defined by observation of $g_{KO}(r)$ only. In particular, the first
minimum of $g_{KO}(r)$ is shallow, and the principal maximum shows a
distinct shoulder that delineates a second population of two water molecules
beyond the inner-most four water molecules.  That second set overlaps an
anticipated first minimum region. These features are distinctly
different for the corresponding results for Li$^+$(aq) and, to a lesser
degree, Na$^+$(aq).

\section*{Acknowledgements} 
Sandia is a multiprogram laboratory operated by Sandia Corporation, a
Lockheed Martin Company, for the US Department of Energy's National
Nuclear Security Administration under contract DE-AC04-94AL85000. The
work at Los Alamos was supported by the US Department of Energy,
contract W-7405-ENG-36, under the LDRD program at Los Alamos.
LA-UR-03-8005.

%The references should start on their own page.
\clearpage

%\begin{thebibliography}{99}
%\bibitem{marker}
%Author(s), {\it Journal title}, Year, {\bf Volume}(Issue number), First page number.
%% BibTeX users can use a style-file for PCCP, which can be found on the
%CTAN internet site of LaTeX:
%http://www.tex.ac.uk/tex-archive/biblio/bibtex/contrib/chem-journal/pccp
%.bst
%\end{thebibliography}

%\bibliography{metals}
%\bibliography{metals,metals_SBR}

\clearpage
%The tables should be submitted normally after the reference list, starting on a separate page.
%\begin{table}
%\begin{center}
%\caption{\label{tab1}Insert table caption here}
%\begin{tabular}{llll}
%\hline
%Header 1 & Header 2 & \multicolumn{2}{l}{Header 3}\\ \cline{3-4}
%&&Subheader 1 & Subheader 2 \\ \hline
%Column 1 & Column 2 & Column 3 & Column 4\\
%Column 1 & Column 2 & Column 3 & Column 4\\
%\hline
%\end{tabular}
%\end{center}
%\end{table}

\begin{figure} 
\begin{center} 
\includegraphics[width=6.5in]{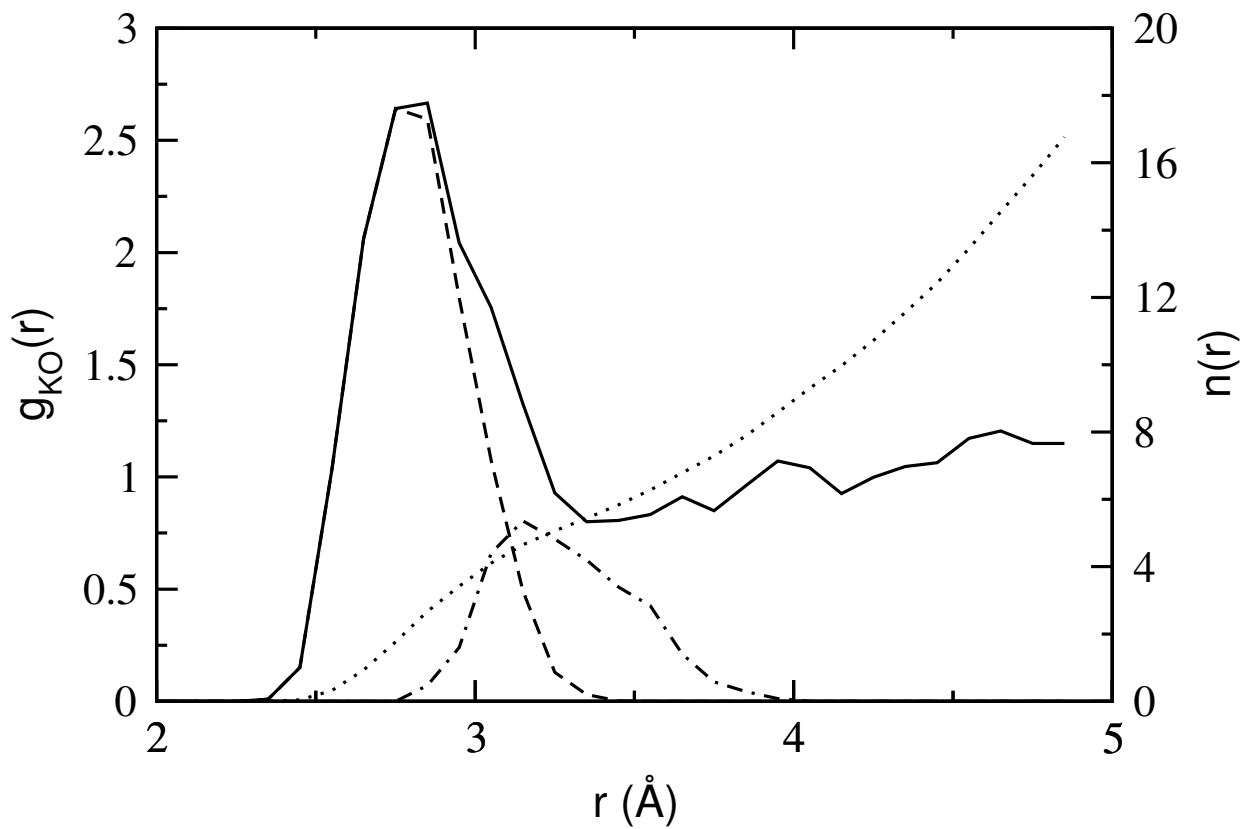} 
\end{center} 
\caption{Oxygen radial density distribution around K$^+$(aq) at infinite
dilution from AIMD simulations. The contribution of the first four water
molecules to the density distribution is shown by the dashed line.
The dot-dashed line shows the contribution of the next two water
molecules to the density distribution. The composite radial distribution
function reflects these two populations as a distinct shoulder on the
outside of the principal maximum. The monotonically increasing dotted
curve is the mean oxygen occupancy of a sphere of radius $r$ centered on
the metal ion, associated with the right vertical axis. 
}\label{fg:gk}
\end{figure}

%Please compile a list of all figure captions on a separate page:
\clearpage
\begin{list}{}{\leftmargin 2cm \labelwidth 1.5cm \labelsep 0.5cm}

\item[\bf Fig. 1] The upper trace (left axis) follows the number of water
molecules within r=3.50~\AA\ of the K$^+$ ion at each time step in the
dynamical simulation.  The radius defines the inner shell of
coordinating water molecules, as determined by the first minimum in the
radial density distribution analysis (Fig.~\ref{fg:gk}). The lower trace
(right axis) records the instantaneous system temperature.  Only the
last 10~ps of this record was used in the structural analysis. 

\item[\bf Fig. 2] Oxygen radial density distribution around K$^+$(aq) at infinite
dilution from AIMD simulations. The contribution of the first four water
molecules to the density distribution is shown by the dashed line.
The dot-dashed line shows the contribution of the next two water
molecules to the density distribution. The composite radial distribution
function reflects these two populations as a distinct shoulder on the
outside of the principal maximum. The monotonically increasing dotted
curve is the mean oxygen occupancy of a sphere of radius $r$ centered on
the metal ion, associated with the right vertical axis.

\end{list}

\clearpage

%\begin{figure}[ht]
%  \begin{center}
%   \includegraphics{filename.eps}
%   \caption{Figure caption goes here.}
%  \end{center}
%\end{figure}

\end{document}